# Infrared bands of CS$_2$ dimer and trimer at 4.5 μm

## A.J. Barclay,[1] K. Esteki,[1] K.H. Michaelian,[2] A.R.W. McKellar,[3]

## N. Moazzen-Ahmadi[1,a]

[1] Department of Physics and Astronomy, University of Calgary, 2500 University Drive North West, Calgary, Alberta T2N 1N4, Canada

[2] Natural Resources Canada, CanmetENERGY, 1 Oil Patch Drive, Suite A202, Devon, Alberta T9G 1A8, Canada

[3] National Research Council of Canada, Ottawa, Ontario K1A 0R6, Canada

[a] Corresponding author. Electronic mail: ahmadi@phas.ucalgary.ca

**Abstract**

We report observation of new infrared bands of $(CS_2)_2$ and $(CS_2)_3$ in the region of the $CS_2$ $\nu_1 + \nu_3$ combination band (at 4.5 μm) using a quantum cascade laser. The complexes are formed in a pulsed supersonic slit-jet expansion of a gas mixture of carbon disulfide in helium. We have previously shown that the most stable isomer of $(CS_2)_2$ is a cross-shaped structure with $D_{2d}$ symmetry and that for $(CS_2)_3$ is a barrel-shaped structure with $D_3$ symmetry. The dimer has one doubly degenerate infrared-active band in the $\nu_1 + \nu_3$ region of the $CS_2$ monomer. This band is observed to have a rather small vibrational shift of -0.846 cm$^{-1}$. We expect one parallel and one perpendicular infrared-active band for the trimer but observe two parallel and one perpendicular bands. Much larger vibrational shifts of -8.953 cm$^{-1}$ for the perpendicular band and -8.845 cm$^{-1}$ and +16.681 cm$^{-1}$ for the parallel bands are observed. Vibrational shifts and possible vibrational assignments, in the case of the parallel bands of the trimer, are discussed using group theoretical arguments.

## 1. Introduction

Weak non-bonding intermolecular forces determine the behavior of gas-phase environments, condensation/solvation processes, physical and chemical properties of bulk matter, and play an important role in biochemistry. These interactions are governed by the electronic structure of the interacting molecules, and are difficult to calculate directly by *ab initio* methods because of the notorious electron correlation problem. To help overcome this problem, synergy between experiment and theory is used to refine the methods to calculate accurate potential energy surfaces (PESs). This synergy has led to development of highly accurate global potential energy surfaces for many dimers, e.g., $(CO_2)_2$,[1] $(OCS)_2$,[2] $(N_2O)_2$,[3,4] $(CS_2)$-$(CO_2)$,[5] and $(CO)_2$.[6,7,8]

In contrast to $CO_2$, $N_2O$ and OCS which form planar slipped-parallel ground state structures with centrosymmetric $C_{2h}$ symmetry [9], the isovalent molecule $CS_2$ forms dimers with three dimensional cross-shaped structures [10,11,12]. The most stable structure of $(CS_2)_2$ has $D_{2d}$ symmetry. As a result, it has one infrared-active fundamental in the region of the $v_3$ fundamental of the $CS_2$ monomer. This vibrational fundamental and a much weaker combination band involving the intermolecular scissor vibration have been observed previously [11]. The appearance of these bands corresponds to those of a perpendicular ($\Delta K = \pm 1$) symmetric rotor molecule band, confirming the cross-shaped structure of $CS_2$ dimer. The vibrational shift for the fundamental and the intermolecular frequency were found to be $-1.221$ cm$^{-1}$ and $10.96(1)$ cm$^{-1}$, respectively. Very recently, Coulomb explosion and laser-based molecular alignment were used to determine the structure of $CS_2$ dimer in helium nanodroplets [13]. $CS_2$ dimer was also found to be 90° cross-shaped, similar to the minimum energy configuration observed in the gas phase.

The most stable structure of $CS_2$ trimer is the symmetric $D_3$ form with three equivalent monomers [14]. The expected infrared-active trimer fundamentals in the region of the $\nu_3$ fundamental of the $CS_2$ monomer are one parallel and one perpendicular band. Both bands have been observed, providing information on the centers of mass distance of the monomers ($r_{C-C}$ = 3.811 Å) and the tilt angle of the monomer axis with the plane containing the C atoms (61.8°). Much larger vibrational shifts compared to that for $CS_2$ dimer were observed, namely +10.31 cm$^{-1}$ for the parallel and −10.74 cm$^{-1}$ for the perpendicular band. Also relevant to the present work is the detection of an isomer of $CS_2$ tetramer having $D_{2d}$ symmetry where the four monomers are equivalent and diagonally opposite monomers are co-planar [15]. Although this isomer is most likely a higher energy structure, it has been identified spectroscopically due to its prominent parallel band having a simple spectral pattern characteristic of a symmetric top with unresolved K structure. This band, which was also observed in the region of the $\nu_3$ fundamental of the $CS_2$ monomer, has a substantial vibrational blue-shift (+16.082 cm$^{-1}$).

For a dimer or larger cluster of identical polyatomic monomers, the possible vibrational states are significantly different in a monomer combination band region than in a fundamental band region. In the present case, the $\nu_1 + \nu_3$ vibration of $CS_2$ monomer occurs at 2185.462 cm$^{-1}$, while the sum of the $\nu_1$ and $\nu_3$ fundamentals is 2193.336 cm$^{-1}$. A cluster has possible vibrations correlating not only with the former ($\nu_1$ and $\nu_3$ on the same monomer) but also with the latter ($v_1$ and $v_3$ on different monomers). In each case, the actual cluster vibrational frequencies may be shifted away from their monomer values by intermolecular interactions, but the two distinct dissociation channels remain. Transition probabilities to states with vibrations on different monomers from the ground vibrational state are likely to be small, since they depend on relatively weak intermolecular coupling. This is explored below in more detail. But these states are still present, and are located relatively nearby, so they may still interact with the more easily observed states (those having vibrations on the same monomer).

In this work, we report the observation of new infrared bands for $(CS_2)_2$ and $(CS_2)_3$ in the region of the $CS_2$ $v_1+ v_3$ combination band (around 4.5 μm). For the dimer we observe a relatively strong perpendicular band centered around 2184.617 cm$^{-1}$. This corresponds to a vibrational red-shift of -0.846 cm$^{-1}$. We show that there are two possible infrared-active bands in this region and conclude that the band observed corresponds to simultaneous excitation of the symmetric stretch ($v_1$) and the asymmetric stretch ($v_3$) on the same monomer. The situation is considerably more complicated for the trimer. We expect five infrared-active bands, two parallel and three perpendicular. We show that only one parallel and one perpendicular band, those with simultaneous excitation $v_1$ and $v_3$ on the same monomer, should show significant infrared activity to be observable. However, we actually observe one perpendicular and two parallel bands, which leaves the possibility that one of the parallel bands could be a perturbation allowed band or a combination band involving a low frequency intermolecular mode. In the following section, dimer and trimer vibrational shifts are discussed using group theoretical arguments.

## 2. Theory

### 2.1. CS₂ dimer

The cross-shaped $CS_2$ dimer with $D_{2d}$ point group symmetry is illustrated in Fig. 1. Here, the *a* inertial axis lies along the line connecting the two carbon atoms, which is also the symmetry axis. There are four possible intramolecular C–S stretching vibrations. They

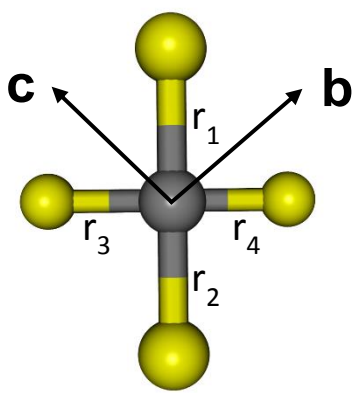

Figure 1: The "cross-shaped" structure of the $CS_2$ dimer and its inertial axis.

can be classified as:

$$\Gamma\left[\left(\Delta r_1 + \Delta r_2\right) + \left(\Delta r_3 + \Delta r_4\right)\right] = \Gamma\left[q_1^a + q_1^b\right] = A_1 , \tag{1}$$

$$\Gamma\left[\left(\Delta r_1 + \Delta r_2\right) - \left(\Delta r_3 + \Delta r_4\right)\right] = \Gamma\left[q_1^a - q_1^b\right] = B_2 , \tag{2}$$

$$\Gamma\left[\left(\Delta r_1 - \Delta r_2\right), \left(\Delta r_3 - \Delta r_4\right)\right] = \Gamma\left[q_3^a, q_3^b\right] = E , \tag{3}$$

where the superscripts *a* and *b* are used to label the monomers. The vibrations with $A_1$ or $B_2$ symmetry correspond to a symmetric stretch within each monomer, which may be mutually in-phase ($A_1$) or out-of phase ($B_2$). Both of these fundamentals occur in the region of the $CS_2$ monomer $\nu_1$ fundamental band ($\sim$660 cm$^{-1}$), with $A_1$ being completely infrared inactive and $B_2$ expected to be very weak since it arises from an induced dipole moment along the dimer $a$-axis. The doubly degenerate vibration $E$ corresponds to the asymmetric C–S stretch within each monomer. It gives rise to a perpendicular band, occurring in the region of the $CS_2$ monomer $\nu_3$ fundamental, whose observation was reported in Ref. [11]. The small vibrational shift of -1.221 cm$^{-1}$ in this case is necessarily due to non-resonant interactions of the monomers because any resonant effects would split the doubly degenerate vibrations into two non-degenerate modes.

The symmetry of the vibrational modes for $CS_2$ dimer in the $\nu_1+\nu_3$ region of the $CS_2$ monomer can be deduced from Eqs. (1)-(3). There are two doubly degenerate modes. One of these is obtained from combination of the vibrations with $A_1$ and $E$ symmetry (i.e., $A_1 \otimes E = E$) and the other from combination of the vibrations with $B_2$ and $E$ symmetry (i.e., $B_2 \otimes E = E$). They are given by

$$\Gamma\left[ q_1^a q_3^a, q_1^b q_3^b \right] = E, \tag{4}$$

$$\Gamma\left[ q_1^a q_3^b, q_3^a q_1^b \right] = E. \tag{5}$$

The vibration in Eq. (4) represents the simultaneous excitation of a symmetric and an asymmetric stretch on the same monomer whereas the mode in Eq. (5) is excitation of a symmetric stretch on one monomer and an asymmetric stretch on the other monomer. The question as to which of the two combination bands has significant infrared intensity can be answered from the vibrational selection rules. Assuming there is no significant induced dipole moment, we have

$$\bar{\mu}(q_1^a, q_3^a, q_1^b, q_3^b) = \bar{\mu}^a(q_1^a, q_3^a) + \bar{\mu}^b(q_1^b, q_3^b) \tag{6}$$

We then consider the terms in the Taylor expansion of the dipole moment function in Eq. (6) which give rise to infrared active bands in the $\nu_1 + \nu_3$ region of the CS$_2$ monomer. The relevant expression for the dipole moment is given by

$$\bar{\mu}(\nu_1 + \nu_3) = \left(\frac{\partial^2 \bar{\mu}^a}{\partial q_1^a \partial q_3^a}\right)_e q_1^a q_3^a + \left(\frac{\partial^2 \bar{\mu}^b}{\partial q_1^b \partial q_3^b}\right)_e q_1^b q_3^b = \bar{\mu}^a(1,3) q_1^a q_3^a + \bar{\mu}^b(1,3) q_1^b q_3^b \tag{7}$$

Application of symmetry conditions for the components of the dipole moment, namely

$\Gamma\left[\mu_a\right] = B_2$, and $\Gamma\left[\mu_b, \mu_c\right] = E$ gives $\mu_a = 0$ and

$$\mu_b(\nu_1 + \nu_3) = \mu_\perp(1,3)\left(q_1^a q_3^a - q_1^b q_3^b\right)$$
$$\mu_c(\nu_1 + \nu_3) = \mu_\perp(1,3)\left(q_1^a q_3^a + q_1^b q_3^b\right) \tag{8}$$

where $\mu_b^a(1,3) = -\mu_b^b(1,3) = \mu_c^a(1,3) = \mu_c^b(1,3) = \mu_\perp(1,3)$.

To the extent that Eq. (6) holds true and considering Eq. (8) together with the two degenerate modes given by Eqs. (4) and (5), we expect only one infrared active perpendicular band in $\nu_1 + \nu_3$ region of the monomer. The mode in Eq. (5) with simultaneous excitation of symmetric and asymmetric stretch on the same monomer is infrared active, whereas the mode represented by Eq. (4) with symmetric stretch excitation on one monomer and the asymmetric stretch excitation on the other is infrared inactive. Any infrared activity for the latter mode arises from an induced dipole moment which is expected to be small.

Vibrational shifts can be calculated from the force field selection rules for the two doubly degenerate modes in Eqs. (4) and (5). The energy matrix is given by

$$H = \begin{bmatrix} a_1b_3 + \beta & h_{12} & h_{13} & h_{14} \\ h_{12} & a_3b_1 + \beta & h_{23} & h_{24} \\ h_{13} & h_{23} & a_{13} + \beta' & h_{34} \\ h_{14} & h_{24} & h_{24} & b_{13} + \beta' \end{bmatrix} \tag{9}$$

Here, $a_1b_3 = a_3b_1$ denotes the energy when $\nu_1$ is on one monomer and $\nu_3$ on the other, $a_{13} = b_{13}$ denotes the energy when $\nu_1 + \nu_3$ is on the same monomer, the off-diagonal elements represent the resonant interactions, and $\beta$ and $\beta'$ are the non-resonant interactions. The resonant interactions must be quartic in $q$ and they must transform as the totally symmetric irreducible representation $A_1$. Such totally symmetric combinations are easily obtained from the application of the projection operator, yielding

$$H_{resonant} \propto q_1^a q_1^b \left[ \left( q_3^b \right)^2 + \left( q_3^a \right)^2 \right]. \tag{10}$$

From Eq. (10), we conclude that $h_{12} = h_{34} = 0$, $h_{13} = h_{24} = 0$ and $h_{14} = h_{23} = \alpha$. These symmetry conditions imply that only a transfer of $\nu_1$ quantum of energy from one monomer to the other is possible and any other energy exchange is symmetry forbidden. The symmetry adapted energy matrix in Eq. (9) then reduces to

$$H = \begin{bmatrix} a_1b_3 + \beta & 0 & 0 & \alpha \\ 0 & a_3b_1 + \beta & \alpha & 0 \\ 0 & \alpha & a_{13} + \beta' & 0 \\ \alpha & 0 & 0 & b_{13} + \beta' \end{bmatrix} \tag{11}$$

Equation (11) has two doubly degenerate eigenvalues given by

$$E_\pm = \tfrac{1}{2}\left( a_1b_3 + a_{13} + \beta + \beta' \right) \pm \tfrac{1}{2}\left[ (a_1b_3 - a_{13} + \beta - \beta')^2 + 4\alpha^2 \right]^{1/2} \tag{12}$$

which is the result we expect.

### 2.2. CS₂ trimer

The CS$_2$ trimer with D$_3$ point group symmetry is illustrated in Fig. 2. Here, two of the principal axes lie in the plane containing the C atoms and the *c* axis is along the three-fold symmetry axis of the complex. There are six possible intramolecular C–S stretching fundamentals. They can be classified as:

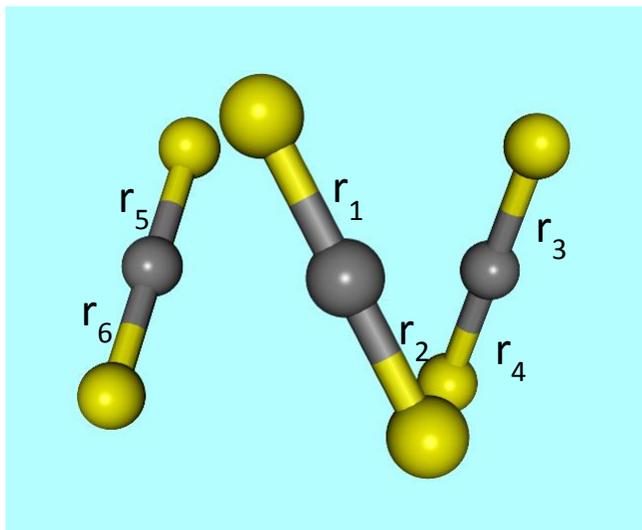

Figure 2: The most stable isomer of CS$_2$ trimer. It has D$_3$ symmetry with the three-fold symmetry axis along the c inertial axis.

$$\Gamma\left[\left(\Delta r_1+\Delta r_2\right)+\left(\Delta r_3+\Delta r_4\right)+\left(\Delta r_5+\Delta r_6\right)\right]=\Gamma\left[q_1^a+q_1^b+q_1^c\right]=A_1 \tag{13}$$

$$\Gamma\left[\left(\Delta r_1-\Delta r_2\right)+\left(\Delta r_3-\Delta r_4\right)+\left(\Delta r_5-\Delta r_6\right)\right]=\Gamma\left[q_3^a+q_3^b+q_3^c\right]=B_2 \tag{14}$$

$$\Gamma\left[2q_1^a-q_1^b-q_1^c,q_1^b-q_1^c\right]=E \tag{15}$$

$$\Gamma\left[2q_3^a-q_3^b-q_3^c,q_3^b-q_3^c\right]=E \tag{16}$$

The vibrations in Eqs. (13) and (15) involve symmetric stretches within each monomer. They are therefore very weak or completely infrared inactive. The remaining two fundamentals, Eqs. (14) and (16), involve the asymmetric C–S stretches within each monomer and are infrared active. Taking the same approach as in section 2.1, we assume that

$$\vec{\mu}(q_1^a, q_3^a, q_1^b, q_3^b, q_1^c, q_3^c) = \vec{\mu}^a(q_1^a, q_3^a) + \vec{\mu}^b(q_1^b, q_3^b) + \vec{\mu}^c(q_1^c, q_3^c) \ . \tag{17}$$

Considering only the relevant terms in the Taylor expansion of the dipole moment, we obtain for the symmetry adapted dipole moment components

$$\mu_z = \mu_\parallel \left( q_3^a + q_3^b + q_3^c \right) \tag{18}$$

$$\mu_x = \frac{1}{2} \mu_\perp \left( 2q_3^a - q_3^b - q_3^c \right) \qquad \mu_y = \frac{\sqrt{3}}{2} \mu_\perp \left( q_3^b - q_3^c \right). \tag{19}$$

Equation (18) gives rise to a parallel band and the dipole moment components in Eq. (19) result in a perpendicular band. Both fundamentals have been observed previously [14]. The energy matrix for the trimer with one quantum of $\nu_3$ excitation is given by

$$H = \begin{bmatrix} a_3 + \beta & h_{12} & h_{13} \\ h_{12} & b_3 + \beta & h_{23} \\ h_{13} & h_{23} & c_3 + \beta \end{bmatrix}$$

Here $h_{12}$ represents an interaction which causes a transfer of $\nu_3$ excitation from monomer $a$ to $b$ and vice versa and $\beta$ is the contribution from all non-resonant interactions. From the transformation properties of the $q$'s, we infer that the lowest order terms that can give rise to an exchange of a $\nu_3$ quantum must be of the form $q_3^a \left( q_3^b \right)^3$. The linear combination of such terms that transforms as totally symmetric irreducible representation $A_1$ is found by application of the projection operator,

$$H_{resonant} \propto \left[ q_3^a \left( q_3^b \right)^3 + q_3^a \left( q_3^c \right)^3 + q_3^b \left( q_3^a \right)^3 + q_3^b \left( q_3^c \right)^3 + q_3^c \left( q_3^a \right)^3 + q_3^c \left( q_3^b \right)^3 \right]$$

resulting in a single resonant interaction parameter. The energy matrix therefore reduces to

$$H = \begin{bmatrix} a_3 + \beta & \alpha & \alpha \\ \alpha & b_3 + \beta & \alpha \\ \alpha & \alpha & c_3 + \beta \end{bmatrix} \qquad (20)$$

Again, this symmetry adapted energy matrix has the correct form since it gives two distinct eigenvalues, a nondegenerate energy, $E_1 = a_3 + 2\alpha + \beta$, and a doubly degenerate energy, $E_2 = a_3 - \alpha + \beta$.

The symmetry of the vibrational modes for $CS_2$ trimer in the $\nu_1+\nu_3$ region of the $CS_2$ monomer is obtained from Eqs. (13)-(16):

$$\left( A_1 + E \right)_{\text{Symmetric stretch}} \otimes \left( A_2 + E \right)_{\text{Antisymmetric stretch}} = A_1 + 2A_2 + 3E . \qquad (21)$$

The vibration with $A_1$ symmetry is infrared inactive whereas the remaining five vibrations are in principle infrared active. The three non-degenerate vibrations can be identified as

$$\Gamma[q_1^a q_3^b + q_1^b q_3^c + q_1^c q_3^a - q_1^b q_3^a - q_1^c q_3^b - q_1^a q_3^c] = A_1 \qquad (22)$$

$$\Gamma[q_1^a q_3^b + q_1^b q_3^c + q_1^c q_3^a + q_1^b q_3^a + q_1^c q_3^b + q_1^a q_3^c] = A_2 \qquad (23)$$

$$\Gamma[q_1^a q_3^a + q_1^b q_3^b + q_1^c q_3^c] = A_2 \qquad (24)$$

The vibration in Eq. (23) can be described as symmetric stretch excitation on one monomer and the asymmetric excitation on another monomer and that in Eq. (24) as simultaneous excitation of the symmetric and antisymmetric stretch on the same monomer. The three doubly degenerate vibrations can similarly be identified as one in which the excitation of $\nu_1$ and $\nu_3$ reside on the same monomer and the other two where the $\nu_1$ excitation is on one monomer and $\nu_3$ excitation is on another monomer.

Using a treatment of the trimer dipole moment function similar to that in section 2.1, we conclude that only two of the five vibrations – the two where the excitation of $\nu_1$ and $\nu_3$ resides on the same monomer – have sufficient infrared activity to be observable, while the remaining vibrations are weak because they arise from an induced dipole moment which is expected to be small. Furthermore, the energy matrix in this case is given by

$$H = \begin{bmatrix} a_{13}+\beta & \alpha_6 & \alpha_6 & \alpha_1 & \alpha_1 & \alpha_3 & \alpha_2 & \alpha_2 & \alpha_3 \\ \alpha_6 & b_{13}+\beta & \alpha_6 & \alpha_2 & \alpha_3 & \alpha_1 & \alpha_1 & \alpha_3 & \alpha_2 \\ \alpha_6 & \alpha_6 & c_{13}+\beta & \alpha_3 & \alpha_2 & \alpha_2 & \alpha_3 & \alpha_1 & \alpha_1 \\ \alpha_1 & \alpha_2 & \alpha_3 & a_3b_1+\beta' & \alpha_5 & \alpha_3 & \alpha_6 & \alpha_3 & \alpha_4 \\ \alpha_1 & \alpha_3 & \alpha_2 & \alpha_5 & a_3c_1+\beta' & \alpha_4 & \alpha_3 & \alpha_6 & \alpha_3 \\ \alpha_3 & \alpha_1 & \alpha_2 & \alpha_3 & \alpha_4 & b_3c_1+\beta' & \alpha_5 & \alpha_3 & \alpha_6 \\ \alpha_2 & \alpha_1 & \alpha_3 & \alpha_6 & \alpha_3 & \alpha_5 & b_3a_1+\beta' & \alpha_4 & \alpha_3 \\ \alpha_2 & \alpha_3 & \alpha_1 & \alpha_3 & \alpha_6 & \alpha_3 & \alpha_4 & c_3a_1+\beta' & \alpha_5 \\ \alpha_3 & \alpha_2 & \alpha_1 & \alpha_4 & \alpha_3 & \alpha_6 & \alpha_3 & \alpha_5 & c_3b_1+\beta' \end{bmatrix}$$

(25)

This matrix has the correct symmetry as it results in six distinct eigenvalues, three of which are doubly degenerate and another three that are non-degenerate.

## 3. Results and analysis

All spectra were obtained using a Daylight Solutions quantum cascade laser to probe a pulsed supersonic slit jet. The expansion gas was a very dilute mixture of about 0.25% $CS_2$ in helium. The backing pressure of about 8 atmospheres resulted in an effective rotational temperature around 2.5 K [16,17]. Wavenumber calibration was made using a fixed etalon and a room temperature $N_2O$ reference gas cell. The PGOPHER computer package was used for spectral simulation and fitting [18].

### 3.1. Dimer spectrum

The dimer band was found near the origin of $\nu_1 + \nu_3$ combination band of the $CS_2$

monomer. As expected, the appearance of the dimer band corresponds to a perpendicular band of a symmetric top rotor. Preliminary analysis of this band was made using the ground state parameters from Ref. [11]. We assigned 122 transitions from this band. These were then combined with the previously observed [11] frequencies for the asymmetric C-S stretching fundamental and the combination band involving this mode plus the scissor intermolecular vibration, in a final frequency analysis involving 288 transitions. The final fit had a root mean square deviation (rms) of $3.2 \times 10^{-4}$ cm$^{-1}$. Molecular parameters arising from the fit are listed in Table I, and a simulation based on these parameters is shown in Fig. 3 along with the observed spectrum.

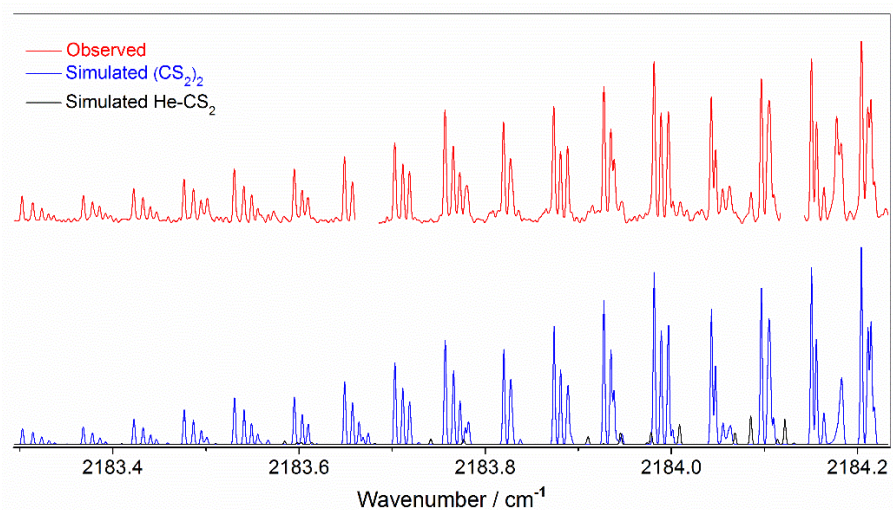

Figure 3: Observed and simulated spectra of the CS$_2$ dimer in the $\nu_1 + \nu_3$ region of the CS$_2$ monomer. The region shown includes a part of the P-branch. The rotational temperature for the simulated spectrum is 2.5 K and the Gaussian line width is 0.002 cm$^{-1}$. Blank regions in the observed trace correspond to strong CS$_2$ monomer lines.

Table I: Molecular parameters for $CS_2$ dimer. Uncertainties in parentheses are 1σ from the least-squares fits in units of the last quoted digit.

| | Ground state | $\nu_3$ band | Combination band | $\nu_1 + \nu_3$ band |
|---|---|---|---|---|
| $\nu_0$/cm$^{-1}$ | | 1534.1347(1) | 1545.0906(2) | 2184.6171(1) |
| $A$ / MHz | 1660.94(13) | 1655.84(14) | 1656.06(18) | 1654.42(14) |
| $B$ / MHz | 802.141(89) | 801.31(10) | 798.07(16) | 801.213(99) |
| $D_J$ / KHz | 2.49(20) | 2.45(28) | 2.99(66) | 2.91(25) |
| $D_K$ / KHz | - | 1.48(65) | - | - |
| $q_+$ / MHz | - | -12.434(44) | -10.880(95) | 0.769(62) |
| $q_-$ / MHz | - | -13.634(44) | -10.10(19) | -1.231(34) |
| $\zeta$ | | - | 0.03760(14) | 0.01047(12) |
| $\Delta\nu$ / cm$^{-1}$ [a] | | -1.221 | | -0.844 |

[a] $\Delta\nu$ is the vibrational shift of the dimer band origin with respect to the band origin of the $CS_2$ monomer.

### 3.2. Trimer spectra

Four trimer bands were recorded in the 2200 cm$^{-1}$ $\nu_1 + \nu_3$ spectral region. Three of these occur in a narrow spectral range around 2176.5 cm$^{-1}$. A portion of the measured spectrum is shown in the upper trace in Fig. 4. The strongest band in this region is a parallel band of $(CS_2)_3$ whose assignment is straightforward due to a strong Q-branch and clear intensity alternation in the P- and R-branch, very similar in structure to its counterpart in the $\nu_3$ fundamental band region (see Fig. 3 in Ref. [14]). The simulated spectrum for this band is shown in the bottom trace in Fig. 4. Note that the top of the Q-branch in both observed and simulated spectra are clipped due to high magnification of the P- and R-branches.

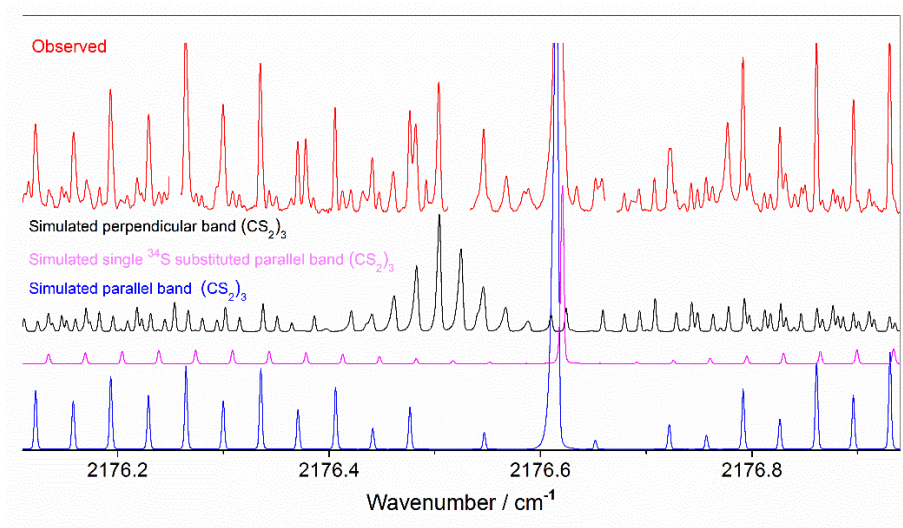

Figure 4: Observed and simulated spectra of the $CS_2$ trimer in the region of the $\nu_1+\nu_3$ band $CS_2$ monomer. The simulated spectrum shows three overlapping bands in the observed spectrum. The rotational temperature for the simulated spectrum is 2.5 K and the Gaussian line width is 0.002 cm⁻¹. Blank regions in the observed trace correspond to strong $CS_2$ monomer lines.

The parallel band of $(CS_2)_3$ is heavily overlapped with a weaker perpendicular trimer band and the absorption lines of $CS_2$ monomer. Assignment of the perpendicular band was possible due to the good simulations that were obtained by keeping the lower state parameters fixed at the values from Ref. [11]. The simulated spectrum for this band is shown in the second trace in Fig. 4. After assigning the lines for these two $CS_2$ trimer bands, we were left with another very weak rotational series with no intensity alternation and a regular but slightly reduced spacing. We concluded that this result is due to the same parallel band, but for trimers containing a single substituted [34]S atom, whose Q-branch is overlapped with that of the parallel band of normal $(CS_2)_3$. Since the natural abundance of [34]S is about 4.2% and there are six equivalent substitution sites, we expect nearly 25% of trimers to contain a single $^{34}$S nucleus. The simulation for this parallel band is illustrated in the third trace in Fig. 4.

The fourth band of the trimer was located around 2202 cm$^{-1}$. A portion of the observed spectrum is illustrated in the top trace in Fig. 5. Although assignment of the band to a parallel band of $(CS_2)_3$ was straightforward due to the existence of a strong central Q-branch (2202.14 cm$^{-1}$) and nuclear spin intensity alternation of the rotational series in the P- and R-branches, this parallel band overlaps with another band with an even stronger Q-branch (2202.18 cm$^{-1}$) surrounded by seemingly complicated rotational structure. All attempts at frequency analysis of this other band, which is possibly due to a larger $CS_2$ cluster, were unsuccessful. The lower trace in Fig. 5 represents the simulated spectrum for the trimer band. Molecular parameters obtained from a simultaneous fit of the three bands reported in this work and those in the region of the $CS_2$ monomer $\nu_3$ fundamental bands are listed in Table II.

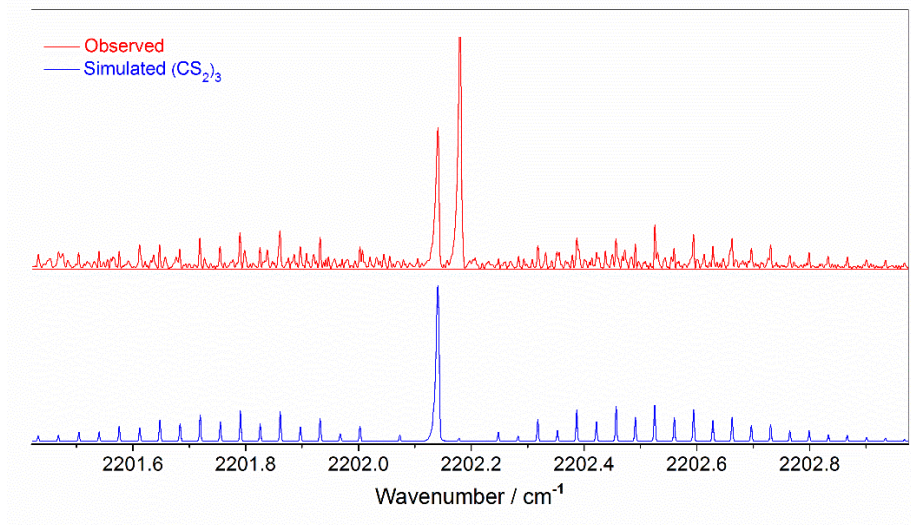

Figure 5: Observed and simulated spectra of the $CS_2$ trimer in the region of the $\nu_1+\nu_3$ band $CS_2$ monomer. This is a parallel band with a strong central Q-branch and clear nuclear spin alternation in the P- and R-branches. The rotational temperature for the simulated spectrum is 2.5 K and the Gaussian line width is 0.002 cm$^{-1}$. The band with the stronger Q branch (2202.18 cm$^{-1}$) has not been analyzed. Blank regions in the observed trace correspond to strong $CS_2$ monomer lines.

Table II: Molecular parameters for $CS_2$ trimer. Uncertainties in parentheses are $1\sigma$ from the least-squares fits in units of the last quoted digit.

|  | Ground state | ⊥ band $\nu_3$ band | ∥ band $\nu_3$ band | ⊥ band $\nu_1+\nu_3$ band | ∥ band $\nu_1+\nu_3$ band | ∥ band $\nu_1+\nu_3$ band |
|---|---|---|---|---|---|---|
| $\nu_0$/ cm$^{-1}$ |  | 1524.6129(1) | 1545.6688(1) | 2176.5083(1) | 2176.6168(1) | 2202.1433(1) |
| $C$ / MHz | 419.201(26) | 419.616(27) | 418.543(32) | 419.501(26) | 418.584(27) | 418.376(28) |
| $B$ / MHz | 525.411(17) | 524.814(19) | 524.759(22) | 524.726(18) | 524.792(18) | 524.584(17) |
| $\zeta \times 10^{-4}$ |  | - | - | 6.1(14) | - |  |
| $\Delta\nu$ / cm$^{-1}$ [a] |  | -10.743 | 10.313 | -8.953 | -8.845 | 16.681 |

[a] $\Delta\nu$ is the vibrational shift of the trimer band origin with respect to the band origin of the $CS_2$ monomer.

### 3. Discussion and Conclusions

The group theoretical discussion in Sec. 2 leads to one infrared active perpendicular band for the dimer in the region of the $CS_2$ monomer $\nu_3$ fundamental band. This vibrational fundamental has been previously observed and analyzed [11], yielding a vibrational shift of -1.221 cm⁻¹. Considering the symmetry adapted energy matrix given by

$$H = \begin{bmatrix} a_3 + \beta & \alpha \\ \alpha & b_3 + \beta \end{bmatrix} \tag{26}$$

where $a_3 = b_3 = 1535.356 \, \text{cm}^{-1}$, and because the eigenvalue of this energy matrix must remain doubly degenerate, we conclude that $\alpha = 0$ and any observed vibrational shift must be due to non-resonant interactions implying that $\beta = -1.221$ cm⁻¹.

We also expect one infrared active perpendicular band for $CS_2$ dimer in the $\nu_1 + \nu_3$ region of the monomer. This is the mode with simultaneous excitation of symmetric and asymmetric stretches on the same monomer represented by Eq. (4). The molecular parameters for the upper state of this band, determined in this work, are given in column 5 of Table I, resulting in a vibrational shift of -0.846 cm⁻¹. However, this vibrational shift is insufficient to fully characterize the energy matrix in Eq. (11), for which two additional pieces of information would be necessary. Note that $a_{13} = b_{13} = 2185.462$ cm⁻¹ and $a_1 b_3 = a_3 b_1 = 2193.336$ cm⁻¹.

The observed vibrational shifts for the trimer bands are given in the last row of Table II. In the $\nu_3$ region of the $CS_2$ monomer these shifts are +10.313 and -10.743 cm⁻¹ for the parallel and perpendicular bands, respectively. These shifts are marginally different from those determined in Ref. [14] where they, together with the shift for the $^{34}$S-substituted parallel band, were used to determine the resonant ($\alpha = -3.75\,\text{cm}^{-1}$) and non-resonant ($\beta = 6.99$ cm$^{-1}$) interactions in Eq. (20).

Analysis of vibrational shifts for the trimer in the $\nu_1 + \nu_3$ region of the CS$_2$ monomer is considerably more complicated. First, from the two modes in Eqs. (23) and (24) with $A_2$ symmetry (parallel band) we expect only the latter to be observable. However, we observe *two* parallel bands (see Table II and Figs. 4 and 5). Therefore, two questions arise: (a) Which of these two bands is the expected "allowed" parallel band; and (b) what is the nature of the unexpected band? It is difficult to be certain but if we choose the band at lower frequency (centered at 2176.617 cm$^{-1}$) to be the expected allowed band, because we observe the corresponding $^{34}$S-substituted parallel band with virtually the same band center, then the other band (centered at 2202.143 cm$^{-1}$) could be a combination band involving a low frequency ($\approx 25$ cm$^{-1}$) intermolecular mode. Another possibility is that the higher frequency band (2202.143 cm$^{-1}$) is due to the mode in which the $\nu_1$ and $\nu_3$ excitations are on different monomers, Eq. 23. Perhaps this mode, which we expected to be weak, gains intensity by mixing with the allowed mode, Eq. (24), via a homogeneous Fermi-type resonance (terms $\alpha_1$, $\alpha_2$, and $\alpha_3$ in Eq. 25). Unfortunately, full experimental characterization of the Eq. (25) energy matrix would require eight vibrational shifts, or other knowledge of parameters, much more than is available at this time.

In conclusion, one infrared band for (CS$_2$)$_2$, three bands for normal species of (CS$_2$)$_3$, and one band for singly substituted $^{34}$S in CS$_2$ trimer in the region of the CS$_2$ $\nu_1 + \nu_3$ combination band (at 4.5 μm) were observed in this work. The observed band for the dimer confirms a 90° cross-shaped structure ($D_{2d}$ symmetry). This band has a relatively small vibrational shift (-0.844 cm$^{-1}$), comparable to its counterpart previously observed in the region of the $CS_2$ monomer $\nu_3$ fundamental (-1.221 cm$^{-1}$). Although the vibrational shift in the latter case can be attributed to non-resonant effects between the monomers, the vibrational shift for the former is insufficient for full characterization of the energy matrix given in Eq. (11).

The vibrational bands observed for $(CS_2)_3$ are all consistent with a twisted barrel-shaped structure ($D_3$ symmetry). We expected to observe two vibrational bands in the 4.5 μm combination band region but found three, one perpendicular and two parallel bands. The unexpected parallel band (centered at 2202.143 cm$^{-1}$) is likely a combination band involving a low frequency intermolecular mode. The vibrational shifts for the trimer bands are significantly larger than those observed for $CS_2$ dimer, but similar in magnitude to the observed shifts for the trimer bands in the region of the $CS_2$ monomer $\nu_3$ fundamental. The vibrational shifts for the $CS_2$ monomer $\nu_3$ fundamental are adequate for full experimental characterization of the energy matrix given in Eq. (20). However, the same cannot be said for the symmetry adapted energy matrix in Eq. (25). This requires eight vibrational shifts or other knowledge of the parameters, much more than the two vibrational shifts determined in this work.

**Acknowledgments**

We gratefully acknowledge the financial support of the Natural Sciences and Engineering Research Council of Canada. We also acknowledge J.T. Hougen for fruitful discussions.

## References

1  X-G. Wang, T. Carrington Jr., R. Dawes, J. Mol. Spectrosc. 330, 179 (2016).

2  J. Brown, X-G. Wang, R. Dawes, T. Carrington Jr., J. Chem. Phys. 136, 134306 (2012).

3  R. Dawes, X-G. Wang, A.W. Jasper, T. Carrington Jr., J. Chem. Phys. 133, 134304 (2010).

4  L. Zheng, Y. Lu, S.Y. Lee, H. Fu, M. Yang, J. Chem. Phys. 134, 054311 (2011).

5  J. Brown, X-G. Wang, T. Carrington Jr., G.S. Grubbs, Richard Dawes, J. Chem. Phys. 140, 114303 (2014).

6  L.A. Surin, I.V. Tarabukin, S. Schlemmer, Y.N. Kalugina, and A. van der Avoird, J. Chem. Phys. 148, 044313 (2018).

7  J-M. Liu, Y. Zhai, X-L. Zhang, H. Li, Phys. Chem. Chem. Phys., 20, 2036 (2018).

8  H. Cybulski, C. Henriksen, R. Dawes, X-G. Wang, N. Bora, G. Avila, T. Carrington, Jr. B. Fernández, Phys. Chem. Chem. Phys., 20, 12624 (2018).

9  N. Moazzen-Ahmadi, A.R.W. McKellar, Int. Rev. Phys. Chem. 32, 611 (2013).

10  C.C. Dutton, D.A. Dows, R. Eikey, S. Evans, R.A Beaudet, J. Phys. Chem. A 102, 6904 (1998).

11  M. Rezaei, J. Norooz Oliaee, N. Moazzen-Ahmadi, A.R.W. McKellar, J. Chem. Phys., 134, 144306 (2011).

12  J. Norooz Oliaee, F. Mivehvar, M. Dehghany, and N. Moazzen-Ahmadi, J. Phys. Chem. A 114, 7311 (2010).

13  J.D. Pickering, B. Shepperson, B.A.K. Hübschmann, F. Thorning, H. Stapelfeldt, Phys. Rev. Lett. 120, 113202 (2018).

14  M. Rezaei, J. Norooz Oliaee, N. Moazzen-Ahmadi and A. R. W. McKellar, Phys. Chem. Phys. Chem. Phys. 13, 12635 (2011).

15  M. Rezaei, J. Norooz Oliaee, N. Moazzen-Ahmadi, A.R.W. McKellar, Chem. Phys.

Lett. 570, 12 (2013).

16  M. Dehghany, Mahin Afshari, Z. Abusara, C. Van Eck, N. Moazzen-Ahmadi, J. Mol.

Spectrosc. 247, 123 (2008).

17  M. Dehghany, Mahin Afshari, R.I. Thompson, N. Moazzen-Ahmadi, A.R.W.

McKellar, J. Mol. Spectrosc. 252, 1 (2008).

18  PGOPHER version 8.0, C. M. Western, 2014, University of Bristol Research Data

Repository, doi:10.5523/bris.huflggvpcuc1zvliqed497r2.